\documentclass[smallextended,runningheads,nospthms]{svjour3}
\smartqed  
\usepackage{amsmath,amsfonts}
\usepackage{amsthm}
\usepackage{graphicx}
\usepackage{textcomp}
\usepackage{subfig,caption}
\usepackage{epsfig}
\usepackage{enumerate}
\theoremstyle{plain}
  \newtheorem{theorem}{Theorem}

\newtheorem{corollary}[theorem]{Corollary}

\title{
The Logarithmic Super Divergence and its use in
Statistical Inference
}
\date{}

\author{
Avijit Maji \and Abhik Ghosh \and Ayanendranath Basu \\
}

\institute{Avijit Maji \and
Abhik Ghosh \and
Ayanendranath Basu
\at Indian Statistical Institute, 203, B.T. Road, Kolkata-700108, India.
\\Tel.: +91 33 2575 2806, Fax: +91 33 2577 3104.
\\\email{avijit.maji$@$hotmail.com, abhianik@gmail.com, 
ayanbasu$@$isical.ac.in.}
}

\begin{document}
\maketitle

\begin{abstract}
This paper introduces a new superfamily of divergences 
that is similar in spirit to the 
$S$-divergence family introduced by Ghosh et al.~(2013).
This new family 
serves as an umbrella 
that
contains the 
logarithmic power divergence family (Renyi, 1961; Maji, Chakraborty
and Basu 2014) and
the logarithmic density power divergence family (Jones et al., 2001)
as special cases.
Various properties of this new family 
and the corresponding minimum distance 
procedures are discussed with particular emphasis 
on the robustness issue; 
these properties are demonstrated through
simulation studies.
In particular the method demonstrates 
the limitation of the first order influence 
function in assessing the robustness 
of the corresponding minimum distance procedures.
\end{abstract}


\section{Introduction}
\label{Sec_Introduction}

The 
density-based minimum divergence approach,
which includes both $\chi^2$ type (Csis\'{z}ar, 1963) and 
Bregman (Bregman, 1967) divergences,
has long history. 
A prominent member of the class of density-based divergences 
is the Pearson's $\chi^2$ (Pearson, 1900)
which started its journey 
from the very early days
of formal research
in statistics. 
From the robustness perspective, however, Beran's 1977 
work is the first useful reference in the literature
of density-based minimum divergence inference. 
In the present paper
we focus on 
a new subclass of density based divergences which encompasses 
some variants of the power divergence measure of 
Cressie and Read (1984) and the density power divergence of 
Basu et al.~(1998)
and discuss 
possible applications in statistical inference. 
Among many other 
things, our analysis highlights the limitation of the first 
order influence function analysis as an indicator of the robustness 
of these procedures. 
\paragraph{}
In this article our primary aim is to describe some statistical uses of the proposed
superfamily of divergences.
To keep this focus 
clear, we will push most of the technical details including the proofs 
of the asymptotic distribution to a separate article, and will 
simply state the relevant theoretical results appropriately in the present context. 
The asymptotic results will be presented in Maji, Ghosh and Basu (2014).  
\paragraph{}
The rest of the paper is organized as follows. 

\section{The Logarithmic Super Divergence and Parametric Estimation}
\label{SEC:lsd_defn}
We first define the generalized $S$-divergence (GSD) family.
Given two probability density functions $g$ and $f$ 
with respect to the same measure, the GSD family is defined,
as a function of two real parameters $\beta$ and $\gamma$, as 
\begin{eqnarray}
\label{EQ:gsd_form}
		{\rm GSD}_{\beta, \gamma}(g,f) 
 &=&  \frac{1}{A} ~ \psi\left(\int f^{1+\beta}\right)  -   \frac{1+\beta}{A B} ~ 
\psi\left(\int f^{B} g^{A}\right)  + \frac{1}{B} ~ \psi\left(\int g^{1+\beta}\right), \nonumber \\
& & ~~~ \beta>0, -\infty<\gamma<\infty,
\end{eqnarray}
where
$A = 1+\gamma (1-\beta)$ and $B = \beta - \gamma (1-\beta)$, 
and $\psi(x)$ is a function with suitable properties. Note that 
$\psi(x)=x$ in (\ref{EQ:gsd_form}) 
recovers
the $S$-divergence family considered by Ghosh et al. (2013);
the function
$\psi(x)=\log x$ 
generates another family of divergences 
which we will refer to as the 
logarithmic super divergence 
(logarithmic $S$-divergence or LSD for short).
The generalizaion given in (\ref{EQ:gsd_form})
is in the spirit of the general form 
considered by Kumar and Basu (2014) in relation to 
the density power divergence measure. However we will
defer the exploration of the properties of this 
generalized divergence 
(including the properties that $\psi$ must possess to be 
statistically useful)
to a sequel paper, and 
concentrate on the properties of the LSD family 
in the present paper. 
The
Logarithmic $S$-Divergence (LSD) 
has the form
\begin{eqnarray}
\label{lsd_form}
		{\rm LSD}_{\beta, \gamma}(g,f) 
 &=&  \frac{1}{A} ~ \log \int f^{1+\beta}  -   \frac{1+\beta}{A B} ~ 
\log \int f^{B} g^{A}  + \frac{1}{B} ~ \log \int g^{1+\beta},
\end{eqnarray}
where
$A$ and $B$ are as defined earlier.
It has to be noted that, $ A+B=1+\beta $.
For $\beta=0~(A=1+\gamma,B=-\gamma),$ this family 
coincides with the 
logarithmic power divergence (LPD) family 
with parameter $\gamma$ 
where LPD has the form 
\begin{equation}
\label{EQ:LPD}
{\rm LPD}_{\gamma}(g,f)=\frac{1}{\gamma(\gamma+1)} \log \int 
\frac{g^{1+\gamma}}{f^\gamma} , \gamma \in \mathbb{R},
\end{equation}
while $\gamma=0$ gives the logarithmic density power divergence (LDPD) family
with parameter $\beta$ 
where LDPD has the form 
\begin{equation}\label{EQ:LDPD}
{\rm LDPD}_\beta(g,f) = \displaystyle \log \int  f^{1+\beta} - 
\left(1 + \frac{1}{\beta}\right)  
\log \int f^\beta g + \frac{1}{\beta} \log \int g^{1+\beta}, \beta\geq0.
\end{equation}
Clearly, for $\beta=0$ and $\gamma=0$, this family coincides with the 
likelihood disparity (LD) where LD has the form 
\begin{equation}
 {\rm LD}(g,f)=\int g \log \left(\frac{g}{f}\right).
\end{equation}
This is a version of the Kullback-Leibler divergence. On 
the other hand, the value $\beta=1$ generates the 
divergence 
\begin{equation}
 \log \left[\frac{\int f^2 \int g^2}{\{\int fg\}^2}\right] 
\end{equation}
irrespective of the value of $\gamma$. Jones et al. (2001) 
have presented a comparison of the method based on 
DPD and LDPD, where a (weak) preference for DPD was 
indicated. Later on Fujisawa and Eguchi (2008) 
and Eguchi (2013) have reported some advantages for LDPD 
for parameter estimation under heavy contamination. Similar 
comparison between the $S$-divergence and the logarithmic 
$S$-divergence remain among our agenda for future work. 
\begin{theorem} 
	Given two densities $g$ and $f$, the measure 
${\rm LSD}_{\beta, \gamma}(g,f) $ represents a genuine 
statistical divergence for all $\beta \ge 0$ and $\gamma \in \mathbb{R}$.

\begin{proof}

 A simple application of Holder's inequality establishes the above result. 
\end{proof}

\end{theorem}
\subsection{Estimating Equation of the LSD}
\label{SEC:lsd_est_eqn}
Consider a parametric class of model densities 
$\{ f_\theta : \theta \in \Theta \subseteq {\mathbb{R}}^p \}$
and 
suppose that our interest is in estimating 
$\theta$. Let $G$ denote the distribution function corresponding
to the true density $g$. The minimum 
LSD
functional
$T_{\beta,\gamma}(G)$ at $G$ is defined through the relation 
\begin{equation}
 {\rm LSD}_{\beta, \gamma}\left(g,f_{T_{\beta,\gamma}(G)}\right)
= \min \limits_{\theta \in \Theta } {\rm LSD}_{\beta, \gamma}(g,f_\theta).
\end{equation}
%
A simple differentiation gives us the estimating equation 
for $\theta$, which is 
\begin{equation}
\label{EQ:lsd_est_eqn}
 \frac{\int f_\theta^{1+\beta}  u_\theta}{\int f_\theta^{1+\beta}} = 
\frac{\int f_\theta^{B} g^A  u_\theta}{\int f_\theta^{B} g^A }.
\end{equation}
For $\beta=0~(A=1+\gamma,B=-\gamma)$, the equation becomes the same as the estimating 
equation of the logarithmic power divergence family with parameter $\gamma$. 
For $\gamma=0~(A=1, B=\beta)$, on the other hand, it is the estimating 
equation for the LDPD measure. 
It takes the value $\theta$ when the true density $g = f_\theta$ is in the model; when 
it does not, $\theta_{\beta,\gamma}^g = T_{\beta,\gamma}(G)$ represents 
the best fitting parameter, and $f_{\theta^g}$ is the model element closest 
to $g$ in terms of logarithmic super divergence. For simplicity in 
the notation, we suppress the 
scripts and refer to $\theta_{\beta,\gamma}^g$ as simply $\theta$ 
when there is no scope for confusion. 

\subsection{Influence Function}\label{SEC:IF_MLSDE} 
\begin{figure}
\centering
\includegraphics[width=130mm, height=100mm] {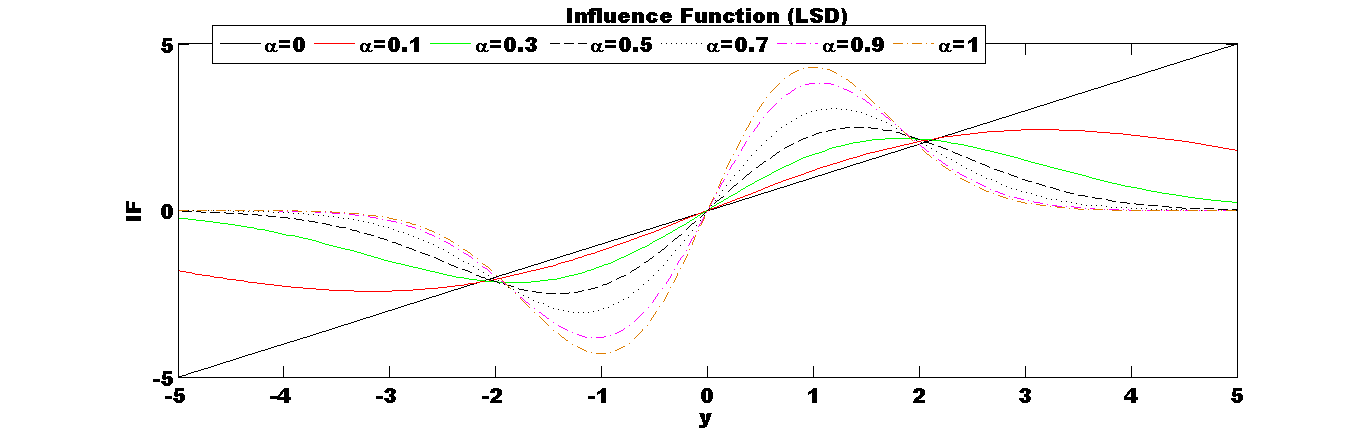}
\caption{Influence function for the $N(\theta,1)$ Distribution at the model with $\theta=0$}
\label{fig:lsd_if_colour}
\end{figure}

The influence function is one of the most important 
heuristic tools in robust inference. 
Consider the minimum LSD functional $T_{\beta,\gamma}(G)$. The 
value $\theta=T_{\beta,\gamma}(G)$ solves the equation (\ref{EQ:lsd_est_eqn}).
Consider the estimating equation 
at the mixture contamination 
density 
$g_\epsilon(x) = (1-\epsilon)~g(x)+\epsilon~I_y(x)$ where $I_y(x)$ is the indicator function at $y$. 
Let $\theta_\epsilon$ be the corresponding functional which solves the estimating 
equation in this case. Taking a derivative of both sides of this estimating 
equation and evaluating at $\epsilon=0$, 
the 
influence function is found to be 
\begin{equation}
IF(y,T,G)=A J_g^{-1}(\theta) {b}(\theta),
\end{equation}
where 
$\theta=T_{\beta,\gamma}(G)$,
\begin{eqnarray}
\label{EQ:j_functional_form}
 J_g(\theta)&=&(1+\beta) \int f_\theta^{1+\beta} u_\theta u_\theta^T 
\int f_\theta^B g^A -\int f_\theta^{1+\beta} i_\theta \int f_\theta^B g^A \nonumber \\
& &
-A \int f_\theta^{1+\beta}u_\theta \int f_\theta^B g^A u_\theta 
-B \int f_\theta^{1+\beta} \int f_\theta^B g^A u_\theta u_\theta^T  \nonumber \\
& &
+ \int f_\theta^{1+\beta} \int f_\theta^B g^A i_\theta, 
\end{eqnarray}    
\begin{eqnarray}
     {b}(\theta) &=& \left(\int f_\theta^{1+\beta}u_\theta~\int f_\theta^B g^A - 
f_\theta^B(y) g^{A-1}(y)~\int f_\theta^{1+\beta}u_\theta \right) \nonumber \\
& &-\left(\int f_\theta^{1+\beta} \int f_\theta^B g^A u_\theta - 
f_\theta^B(y) g^{A-1}(y) u_\theta(y) \int f_\theta^{1+\beta} \right).
\end{eqnarray}
In the above $i_\theta=-\nabla u_\theta$, where $\nabla$
represents the gradient with respect to $\theta$. 
When the model holds, so that $g=f_\theta$ for some $\theta$, 
the influence function becomes,
\begin{eqnarray}
\label{EQ:IF_0}
 IF(y,T,F_\theta)&=& J_0(\theta)^{-1} \left({f_\theta^\beta(y)~\left[u_\theta(y) \int f_\theta^{1+\beta} 
-\int f_\theta^{1+\beta} u_\theta\right]}\right),
\end{eqnarray}
where
\begin{eqnarray}
\label{EQ:j_0_form}
J_0(\theta) = \left({\int f_\theta^{1+\beta}
u_\theta u_\theta^T ~\int f_\theta^{1+\beta} -
\left[\int f_\theta^{1+\beta}~u_\theta\right] 
\left[\int f_\theta^{1+\beta}~u_\theta\right]^T}\right).
\end{eqnarray}
When $\beta=0$, $J_0(\theta)$ reduces to $I(\theta)$, the Fisher information. 
The remarkable observation in (\ref{EQ:IF_0}) and (\ref{EQ:j_0_form}) is that 
the influence function at the model 
is independent of $\gamma$ and depends only on $\beta$. From Figure \ref{fig:lsd_if_colour}
it is clear that the first order influence function is unbounded 
for $\beta=0$ whereas for other values of $\beta$ the function 
is 
bounded and redescending. We will demonstrate the limitations 
of this measure in our context in the subsequent sections. 


\section{Asymptotic Distribution of the Minimum LSD Estimators in Discrete Models}

Under the parametric set-up of Section {\ref{SEC:lsd_est_eqn}},
consider a discrete family of distributions. We will use 
the term ``density function'' generally for the sake of a 
unified notation, irrespective of whether the distribution is discrete
or continuous. Let $X_1,\ldots,X_n$ be a random sample from the 
true distribution having density function $g$ and
let the distribution have support $\chi = \{ 0,1,2,\cdots \}$. Denote the 
relative frequency at $x$ from the data 
by $r_n(x) = \frac{1}{n} \sum_{i=1}^n 
I_x(X_i)$.
Representing 
the logarithmic $S$-divergence in terms of the parameter $\beta$ and $\gamma$
(as given in Section \ref{SEC:lsd_defn}), let 
$\hat{\theta}_{\beta,\gamma}$ be the estimator obtained by minimizing 
${\rm LSD}_{\beta,\gamma}(\hat{g},f_\theta)$ over $\theta \in \Theta$, where 
$\hat{g}$ is a suitable nonparametric density estimate of $g$; 
in the discrete case the vector of relative frequencies $r_n$
based on the sample data is the canonical choice for $\hat{g}$. 
\paragraph{}
In this paper we will primarily describe the statistical applications 
of the minimum distance procedures that are generated by the 
logarithmic $S$-divergence. However, for the sake of completeness,
we also present the asymptotic distribution of the estimators 
which has been separately established in 
Maji, Ghosh and Basu (2014).
\paragraph{}
When $g$ is replaced by $r_n$, the  
estimating equation (\ref{EQ:lsd_est_eqn}) 
may be expressed as 
\begin{eqnarray}
    \sum_x M(\delta_n(x)) f_{\theta}^{1+\beta}(x) w_{\theta}(x) = 0,
\end{eqnarray}
where $$w_{\theta}(x)=[B(\theta)u_\theta(x)-A(\theta)],$$
$$M(\delta)=\delta^A-1, A(\theta)=\sum_x 
f_{\theta}^{1+\beta}(x)u_{\theta}(x), 
B(\theta)=\sum_x 
f_{\theta}^{1+\beta}(x),$$ 
$\delta_n(x) = \frac{r_n(x)}{f_{\theta}(x)}$
and $\delta_g^g(x) = \frac{g(x)}{f_{\theta^g}(x)}$.

\noindent
Define,
\begin{eqnarray}
\label{k_j_exp}
J_g &=& E_g \left[ w_{\theta^g}(X)u_{\theta^g}^T(X) M'(\delta_g^g(X))
 f_{\theta^g}^{\beta}(X) \right] 
-  \sum_x  M(\delta_g^g(x)) 
f_{\theta}^{1+\beta}(x) \nabla w_{\theta^g}(x) \nonumber \\
 & & 
- (1+\beta) \sum_x  M(\delta_g^g(x)) 
f_{\theta}^{1+\beta}(x) w_{\theta^g}(x) u_{\theta^g}(x)  
\end{eqnarray}
and
\begin{eqnarray}
\label{EQ:k_form}
    K_g &=& Var_g \left[ M'(\delta_g^g(X)) 
f_{\theta^g}^\beta(X) w_{\theta^g}(X) \right].
\end{eqnarray}
Note that the matrices $J_g$ in (\ref{EQ:j_functional_form}) 
and (\ref{k_j_exp}) are identical.   
Then, under standard regularity conditions 
(See 
Maji, Ghosh and Basu, 2014),
it 
follows 
that $\hat{\theta}_{\beta,\gamma}$
is consistent for ${\theta}$ and 
has the asymptotic distribution given by
\begin{equation}
n^{\frac{1}{2}}(\hat{\theta}_{\beta,\gamma}-
\theta)\rightarrow N(0,J_g^{-1}K_g J_g^{-1}),
\end{equation}
as $J_g$ and $K_g$ are as defined in (\ref{k_j_exp})
and \ref{EQ:k_form}.
See Maji, Ghosh and Basu (2014) for the technical details of the proof.
\begin{corollary}
	When the true distribution $G$ belongs to the model family, 
i.e., $G = F_\theta$ for some $\theta \in \Theta$, then 
$n^{1/2} (\theta_n - \theta)$ has asymptotic distribution 
as $N_p( 0, J^{-1} V J^{-1} )$, where 
\begin{eqnarray}
     J = J_\beta(\theta) &=&  E_g[w_\theta(X)u_\theta(X)^T 
f_\theta^{\beta}(X)] \nonumber \\ &=& \sum_x \{B(\theta) u_\theta(x)-A(\theta)\}
u_\theta^T(x) f_\theta^{1+\beta}(x).
\\
	 K = K_\beta(\theta) &=& V_g[w_\theta(X)
f_\theta^{\beta}(X)] \nonumber \\ &=& \sum_x
\{B(\theta) u_\theta(x)-A(\theta)\}  \{B(\theta) u_\theta(x)-A(\theta)\}^T
f_\theta^{1+2\beta}(x) \nonumber \\ & & - \xi \xi^T,
\\
     \xi = \xi_\beta(\theta) &=& E_g[w_\theta(X)
f_\theta^{\beta}(X)] = \sum_x \{B(\theta) u_\theta(x)-A(\theta)\} 
f_\theta^{1+\beta}(x).
\\\nonumber
\end{eqnarray}
%
\end{corollary}
Note that, under model ($g=f_\theta$) both $J$ and $K$ depend only on $\beta$.
Thus, the asymptotic distribution of the minimum LSD estimators do not depend 
on the parameter $\gamma$. 

\section{Testing Parametric Hypothesis using the LSD Measures}

\subsection{One Sample problem}
	We consider a parametric family of densities 
$\mathcal{F} = \{f_\theta : \theta \in \Theta \subseteq \mathbb{R}^p\}$ 
as 
introduced earlier. Suppose we are given a random sample $X_1, \ldots, X_n$ of 
size $n$ from the population. Based on this sample, we want to test the hypothesis 
$$
 H_0 : \theta = \theta_0 ~~~ \mbox{against} ~~~~ H_1 : \theta \ne \theta_0.
$$
When the model is correctly 
specified and the null hypothesis is correct, $f_{\theta_0}$ is the data 
generating density. 
We consider the 
test statistics based on the LSD with parameter 
$\beta$ and $\gamma$ defined by 
\begin{equation}
\label{EQ:w_define}
 W_{\beta,\gamma}(\hat{\theta}_{\beta,\gamma},\theta_0)
=2n~{\rm LSD}_{\beta,\gamma}(f_{\hat{\theta}_{\beta,\gamma}},f_{\theta_0}),
\end{equation}
where ${\rm LSD}_{\beta,\gamma}(f_{\hat{\theta}_{\beta,\gamma}},f_{\theta_0})$ has the form 
given in (\ref{lsd_form}). Then the following theorem 
becomes useful in obtaining
the critical values of 
the test statistics in (\ref{EQ:w_define}). 

\begin{theorem}
 The 
asymptotic distribution of the test statistic 	$W_{\beta,\gamma}
(f_{{\hat{\theta}}_{\beta,\gamma}},f_{{\theta_0}})$, under the null hypothesis 
$H_0 : \theta = \theta_0$, coincides with the distribution of 
$$
\sum_{i=1}^r ~  \zeta_i^{\beta}(\theta_0)Z_i^2
$$
where $Z_1, \ldots,Z_r$ are independent standard normal variables,
$\zeta_1^{\beta}(\theta_0), \ldots, \\ \zeta_r^{\beta}
(\theta_0)$ are the nonzero eigenvalues of $A_{\beta}(\theta_0) 
J_{\beta}^{-1}(\theta_0) K_{\beta}(\theta_0) J_{\beta}^{-1}(\theta_0)$, 
with $J_{\beta}(\cdot)$ and $K_{\beta}(\cdot)$ as defined in 
(\ref{k_j_exp})
and the matrix $A_{\beta}(\theta_0)$ is defined as
$$
A_{\beta}(\theta_0) = \nabla [  \nabla {\rm LSD}_{\beta,\gamma}(f_\theta, f_{\theta_0}) ]
|_{\theta = \theta_0} 
$$
 and
$$
r = rank\left( J_\beta^{-1}(\theta_0) 
K_\beta(\theta_0) J_\beta^{-1}(\theta_0)
A_\beta(\theta_0) J_\beta^{-1}(\theta_0) 
K_\beta(\theta_0) J_\beta^{-1}(\theta_0)\right).
$$
Here $\nabla$ represents the gradient with respect to $\theta$. 
\end{theorem}

\begin{figure}
\centering
\includegraphics[width=130mm]{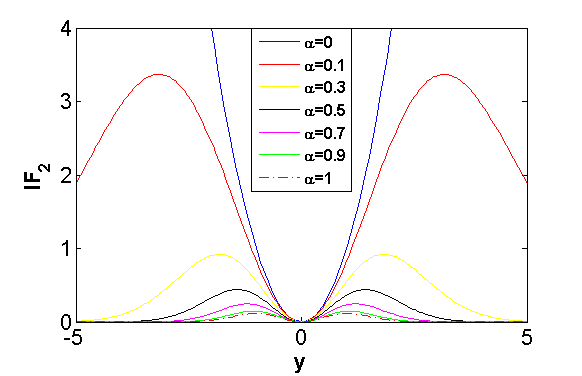}
\caption{Influence function of the LSD test statistics of normal mean at the null $H_0 : \theta=0$}
\label{fig:lsdT_if}
\end{figure}

To see the robustness properties of the LSD based test, we study the influence function analysis 
of the test statistics as in Hampel et al.~(1986), Ghosh and Basu (2014) etc.
We define the corresponding LSD based test functional (LSDT) for one sample simple hypothesis 
problem as described above as (ignoring the sample size dependent multiplier) 
$$
T_{\beta,\gamma}^{(1)}(G) = {\rm LSD}_{\beta, \gamma}\left(f_{T_{\beta,\gamma}(G)},f_{\theta_0}\right),
$$ 
where $T_{\beta,\gamma}(G)$ is the minimum LSD functional defined in Section \ref{SEC:IF_MLSDE}. 
Then, considering the contaminated distribution $G_\epsilon$ associated with $g_\epsilon$, 
Hampel's first-order influence function of the LSDT functional turns out to be zero 
at the null distribution $G=F_{\theta_0}$.  
However, corresponding second order influence function of the LSDT functional at the null distribution
has a non-zero form given by 
\begin{equation}
IF_2(y; T_{\beta,\gamma}^{(1)}, F_{\theta_0}) 
= IF(y; T_{\beta,\gamma}, F_{\theta_0})^T A_{\beta}(\theta_0)  IF(y; T_{\beta,\gamma}, F_{\theta_0}).
\end{equation}
Therefore the robustness of the LSDT functional depends directly on the robustness of the 
minimum LSD estimator used in constructing the test statistics. So, following the arguments of Section 
\ref{SEC:IF_MLSDE} it follows that, the proposed test will have bounded influence function 
whenever $\beta>0$ implying its robustness and has unbounded influence function at $\beta=0$
implying the lack of robustness. Figure \ref{fig:lsdT_if} shows the second order influence function of 
the $N(\theta,1)$ model at the simple null $\theta=0$; the equivalence with the corresponding 
influence function of the minimum LSD estimator presented in Figure \ref{fig:lsd_if_colour} is quite clear.

\subsection{Two Sample Problem}
	Again consider a parametric family of densities 
$\{f_\theta : \theta \in \Theta \subseteq \mathbb{R}^p\}$ as above in 
one sample problem, but here we are given two random 
samples $X_1, \ldots, X_n$ of size $n$ and $Y_1, \ldots, Y_m$ of 
size $m$ from two populations having parameters $\theta_1$ and 
$\theta_2$ respectively and based on these two samples, we want to 
test for the homogeneity of the two samples, i.e. to test the hypothesis 
$$
 H_0 : \theta_1 = \theta_2 ~~~ \mbox{against} ~~~~ H_1 : \theta_1 \ne \theta_2.
$$
 We will consider the estimator $^{(1)}\hat{\theta}_{\beta,\gamma}$  and 
$^{(2)}\hat{\theta}_{\beta,\gamma}$ of $\theta_1$ and $\theta_2$ respectively, 
obtained by minimizing the LSD having parameter ${\beta,\gamma}$ and then as before, we 
consider the test statistic based on the LSD with parameter $\beta$ 
and $\gamma$ as follows
\begin{eqnarray}
    S_{\beta,\gamma}\left( ^{(1)}\hat{\theta}_{\beta,\gamma} , 
^{(2)}\hat{\theta}_{\beta,\gamma}\right) 
= \frac{2nm}{n+m} ~  {\rm LSD}_{\beta,\gamma}\left( f_{^{(1)}\hat{\theta}_{\beta,\gamma}} , 
f_{^{(2)}\hat{\theta}_{\beta,\gamma}}\right).
\end{eqnarray}
We present the asymptotic distribution of the test statistics \\ $S_{\beta,\gamma}
\left( ^{(1)}\hat{\theta}_{\beta,\gamma}, ^{(2)}\hat{\theta}_{\beta,\gamma}\right)$ under $H_0$ 
in the following theorem.

\begin{theorem}
\label{th_two}
The asymptotic distribution of the test 
statistic \\	$S_{\beta,\gamma}\left( ^{(1)}\hat{\theta}_{\beta,\gamma} , 
^{(2)}\hat{\theta}_{\beta,\gamma}\right)$, under the null hypothesis $H_0 : \theta_1 = \theta_2$, 
coincides with the distribution of 
$$
\sum_{i=1}^r ~  \zeta_i^{\beta}(\theta_1)Z_i^2
$$
where $Z_1, \ldots,Z_r$ are independent standard normal variables, 
$\zeta_1^{\beta}(\theta_1), \ldots, \\ \zeta_r^{\beta}(\theta_1)$ 
are the nonzero eigenvalues of $A_\beta(\theta_1) J_\beta^{-1}(\theta_1) 
K_\beta(\theta_1) J_\beta^{-1}(\theta_1)$, with \\
$J_\beta(\cdot)$, $K_\beta(\cdot)$ 
and $A_\beta(\cdot)$ as defined in previous section and
$$
r = rank\left( J_\beta^{-1}(\theta_1) 
K_\beta(\theta_1) J_\beta^{-1}(\theta_1)
A_\beta(\theta_1) J_\beta^{-1}(\theta_1) 
K_\beta(\theta_1) J_\beta^{-1}(\theta_1)\right).
$$
\end{theorem}

\section{Numerical Illustrations}

\subsection{Performance of the Minimum LSD Estimator : Simulation in the Poisson Model}

To explore the performance of the proposed minimum LSD estimators, we have done several simulation 
studies under the Poisson model with sample size of $n=50$. We simulate data from a 
Poisson distribution with parameter $\theta = 4$ and compute the empirical bias and
the MSE of the minimum LSD estimators of $\theta$ based on 1000 replications. 
The results obtained are reported in Tables 
\ref{TAB:bias_uncontaminated} and \ref{TAB:MSE_uncontaminated} respectively. 
Clearly both the bias and MSE are quite small for any $(\beta, \gamma)$ combination;
however the MSE increases slightly with $\beta$.

\begin{table}
\caption{\label{TAB:bias_uncontaminated}
The Empirical bias of the minimum LSD estimators for different values of $\beta$ and $\gamma$}
\centering
\resizebox{\textwidth}{!}{ %
\begin{tabular}{r r r r r r r r r} \hline
$\gamma$	&	$\beta=0$	&	$\beta=0.1$	&	$\beta=0.2$	&	$\beta=0.4$	&	$\beta=0.5$	&	$\beta=0.6$	&	$\beta=0.8$	&	$\beta=1$	\\ \hline 
$-1$	&	$-1.104$	&	$-0.336$	&	$-0.186$	&	$-0.055$	&	$-0.035$	&	$-0.013$	&	0.006	&	0.015	\\
$-0.9$	&	$-0.360$	&	$-0.202$	&	$-0.124$	&	$-0.050$	&	$-0.028$	&	$-0.012$	&	$-0.003$	&	0.010	\\
$-0.7$	&	$-0.169$	&	$-0.113$	&	$-0.073$	&	$-0.029$	&	$-0.019$	&	$-0.008$	&	0.002	&	0.006	\\
$-0.5$	&	$-0.095$	&	$-0.065$	&	$-0.044$	&	$-0.016$	&	$-0.004$	&	$-0.004$	&	0.008	&	0.016	\\
$-0.3$	&	$-0.049$	&	$-0.030$	&	$-0.020$	&	$-0.003$	&	$-0.001$	&	$-0.002$	&	0.004	&	0.020	\\
$-0.1$	&	$-0.015$	&	$-0.010$	&	$-0.002$	&	0.006	&	0.011	&	0.012	&	0.014	&	0.015	\\
0	&	0.000	&	0.000	&	0.009	&	0.017	&	0.012	&	0.004	&	0.007	&	0.024	\\
0.1	&	0.014	&	0.013	&	0.017	&	0.018	&	0.004	&	0.012	&	0.014	&	0.007	\\
0.3	&	0.038	&	0.037	&	0.031	&	0.023	&	0.021	&	0.019	&	0.014	&	0.007	\\
0.5	&	0.060	&	0.053	&	0.046	&	0.030	&	0.021	&	0.026	&	0.019	&	0.006	\\
0.7	&	0.080	&	0.069	&	0.060	&	0.039	&	0.042	&	0.024	&	0.020	&	0.012	\\
0.9	&	0.098	&	0.085	&	0.071	&	0.048	&	0.038	&	0.025	&	0.019	&	0.009	\\
1	&	0.106	&	0.090	&	0.077	&	0.047	&	0.031	&	0.031	&	0.013	&	0.017	\\
1.5	&	0.140	&	0.125	&	0.108	&	0.069	&	0.056	&	0.023	&	0.023	&	0.008	\\
2	&	0.166	&	0.150	&	0.130	&	0.087	&	0.067	&	0.050	&	0.025	&	0.006	\\
\hline
\end{tabular}}
\end{table}
\begin{table}
\caption{\label{TAB:MSE_uncontaminated}
The Empirical MSE of the minimum LSD estimators for different values of $\beta$ and $\gamma$}
\centering
\resizebox{\textwidth}{!}{ %
\begin{tabular}{r r r r r r r r r} \hline
$\gamma$	&	$\beta=0$	&	$\beta=0.1$	&	$\beta=0.2$	&	$\beta=0.4$	&	$\beta=0.5$	&	$\beta=0.6$	&	$\beta=0.8$	&	$\beta=1$	\\ \hline 
$-1$	&	6.989	&	0.415	&	0.251	&	0.136	&	0.147	&	0.131	&	0.142	&	0.148	\\
$-0.9$	&	0.316	&	0.179	&	0.144	&	0.131	&	0.124	&	0.131	&	0.142	&	0.154	\\
$-0.7$	&	0.137	&	0.124	&	0.116	&	0.120	&	0.129	&	0.129	&	0.141	&	0.140	\\
$-0.5$	&	0.101	&	0.101	&	0.104	&	0.115	&	0.122	&	0.122	&	0.140	&	0.153	\\
$-0.3$	&	0.088	&	0.091	&	0.094	&	0.107	&	0.117	&	0.120	&	0.138	&	0.152	\\
$-0.1$	&	0.083	&	0.090	&	0.097	&	0.107	&	0.114	&	0.119	&	0.143	&	0.154	\\
0	&	0.083	&	0.085	&	0.096	&	0.108	&	0.110	&	0.122	&	0.136	&	0.155	\\
0.1	&	0.082	&	0.088	&	0.092	&	0.106	&	0.115	&	0.120	&	0.134	&	0.150	\\
0.3	&	0.084	&	0.086	&	0.094	&	0.106	&	0.115	&	0.122	&	0.133	&	0.148	\\
0.5	&	0.087	&	0.088	&	0.092	&	0.102	&	0.112	&	0.120	&	0.139	&	0.151	\\
0.7	&	0.092	&	0.091	&	0.093	&	0.102	&	0.112	&	0.117	&	0.128	&	0.147	\\
0.9	&	0.099	&	0.096	&	0.095	&	0.103	&	0.111	&	0.112	&	0.129	&	0.150	\\
1	&	0.102	&	0.096	&	0.093	&	0.100	&	0.105	&	0.117	&	0.130	&	0.153	\\
1.5	&	0.121	&	0.111	&	0.104	&	0.103	&	0.105	&	0.109	&	0.128	&	0.153	\\
2	&	0.139	&	0.127	&	0.114	&	0.102	&	0.103	&	0.113	&	0.121	&	0.152	\\
\hline
\end{tabular}}
\end{table}

Next to study the robustness properties of the minimum LSD estimators we repeat the 
above study,  but introduce a contamination in the simulated samples by replacing 
$10\%$ of it by $Poisson(\theta={\rm 12})$ observations. The corresponding values of the 
empirical bias and MSE, against the target value of $\theta = 3$, are presented in Tables 
\ref{TAB:bias_contaminated} and \ref{TAB:MSE_contaminated} respectively. 
Note that, the minimum LSD estimators are seen to be robust for all $\beta \in [0,1]$ if 
$\gamma < 0$ and for suitably large values of $\beta$ if $\gamma \geq 0$.
However, the estimators corresponding to small $\beta$ close to zero and $\gamma \geq 0$.

\begin{table}
\caption{\label{TAB:bias_contaminated}
The Empirical bias of the minimum LSD estimators under $10\%$ contamination for different values of $\beta$ and $\gamma$}
\centering
\resizebox{\textwidth}{!}{ %
\begin{tabular}{r r r r r r r r r} \hline
$\gamma$	&	$\beta=0$	&	$\beta=0.1$	&	$\beta=0.2$	&	$\beta=0.4$	&	$\beta=0.5$	&	$\beta=0.6$	&	$\beta=0.8$	&	$\beta=1$	\\ \hline 
$-1$	&	$-1.407$	&	$-0.140$	&	$-0.024$	&	0.064	&	0.071	&	0.087	&	0.086	&	0.083	\\
$-0.9$	&	$-0.134$	&	$-0.021$	&	0.027	&	0.077	&	0.081	&	0.081	&	0.084	&	0.079	\\
$-0.7$	&	0.056	&	0.084	&	0.090	&	0.106	&	0.099	&	0.089	&	0.092	&	0.072	\\
$-0.5$	&	0.172	&	0.154	&	0.141	&	0.118	&	0.105	&	0.103	&	0.096	&	0.088	\\
$-0.3$	&	0.314	&	0.244	&	0.202	&	0.151	&	0.123	&	0.104	&	0.094	&	0.083	\\
$-0.1$	&	0.578	&	0.394	&	0.283	&	0.174	&	0.143	&	0.123	&	0.102	&	0.082	\\
0	&	0.800	&	0.519	&	0.347	&	0.192	&	0.160	&	0.136	&	0.091	&	0.082	\\
0.1	&	1.071	&	0.697	&	0.439	&	0.213	&	0.160	&	0.144	&	0.108	&	0.085	\\
0.3	&	1.590	&	1.165	&	0.726	&	0.267	&	0.188	&	0.149	&	0.111	&	0.084	\\
0.5	&	1.965	&	1.604	&	1.147	&	0.368	&	0.237	&	0.161	&	0.106	&	0.077	\\
0.7	&	2.219	&	1.929	&	1.532	&	0.546	&	0.289	&	0.183	&	0.117	&	0.081	\\
0.9	&	2.394	&	2.161	&	1.834	&	0.805	&	0.390	&	0.217	&	0.112	&	0.079	\\
1	&	2.461	&	2.252	&	1.950	&	0.958	&	0.452	&	0.240	&	0.115	&	0.083	\\
1.5	&	2.671	&	2.545	&	2.354	&	1.627	&	0.996	&	0.402	&	0.132	&	0.089	\\
2	&	2.773	&	2.691	&	2.568	&	2.055	&	1.545	&	0.792	&	0.149	&	0.084	\\
\hline
\end{tabular}}
\end{table}
\begin{table}
\caption{\label{TAB:MSE_contaminated}
The Empirical MSE of the minimum LSD estimators under $10\%$ contamination for different values of $\beta$ and $\gamma$}
\centering
\resizebox{\textwidth}{!}{ %
\begin{tabular}{r r r r r r r r r} \hline
$\gamma$	&	$\beta=0$	&	$\beta=0.1$	&	$\beta=0.2$	&	$\beta=0.4$	&	$\beta=0.5$	&	$\beta=0.6$	&	$\beta=0.8$	&	$\beta=1$	\\ \hline 
$-1$	&	7.336	&	0.419	&	0.257	&	0.178	&	0.183	&	0.168	&	0.172	&	0.183	\\
$-0.9$	&	0.303	&	0.207	&	0.196	&	0.166	&	0.187	&	0.165	&	0.178	&	0.183	\\
$-0.7$	&	0.160	&	0.159	&	0.158	&	0.157	&	0.166	&	0.169	&	0.196	&	0.176	\\
$-0.5$	&	0.161	&	0.162	&	0.159	&	0.161	&	0.165	&	0.164	&	0.179	&	0.184	\\
$-0.3$	&	0.216	&	0.184	&	0.174	&	0.162	&	0.158	&	0.161	&	0.171	&	0.185	\\
$-0.1$	&	0.430	&	0.268	&	0.203	&	0.168	&	0.158	&	0.166	&	0.169	&	0.177	\\
0	&	0.732	&	0.369	&	0.238	&	0.167	&	0.169	&	0.168	&	0.167	&	0.182	\\
0.1	&	1.276	&	0.581	&	0.302	&	0.184	&	0.161	&	0.165	&	0.172	&	0.183	\\
0.3	&	2.836	&	1.525	&	0.626	&	0.200	&	0.172	&	0.168	&	0.175	&	0.181	\\
0.5	&	4.343	&	2.909	&	1.492	&	0.251	&	0.198	&	0.169	&	0.170	&	0.184	\\
0.7	&	5.524	&	4.207	&	2.669	&	0.409	&	0.208	&	0.176	&	0.166	&	0.186	\\
0.9	&	6.401	&	5.261	&	3.831	&	0.772	&	0.271	&	0.187	&	0.166	&	0.188	\\
1	&	6.749	&	5.703	&	4.328	&	1.075	&	0.319	&	0.194	&	0.174	&	0.181	\\
1.5	&	7.887	&	7.204	&	6.222	&	3.060	&	1.175	&	0.284	&	0.172	&	0.184	\\
2	&	8.462	&	8.001	&	7.335	&	4.820	&	2.779	&	0.773	&	0.170	&	0.188	\\
\hline
\end{tabular}}
\end{table}

\section{Limitation of the First Order Influence Function \& some Remedies}

The numeral examples and simulation results presented in the previous section clearly shows that 
the robustness of minimum LSD estimators in terms of its bias and MSE under data contamination depends 
on the parameter $\gamma$ for smaller values of $\beta$. However, according to the classical literature, 
its first order influence function suggests that (see Section \ref{SEC:IF_MLSDE}) its 
robustness will be independent of the parameter $\gamma$ for all values of $\beta$.
Thus, the classical approach of robustness measure through the first order influence 
fails in the case of minimum divergence estimation with the logarithmic super divergence family.
Similar limitations of the first order influence functions was also observed by Lindsay (1994) 
and Ghosh et al.~(2013) for the case of power divergence family 
and the $S$-divergence family; accordingly they have proposed some alternative measure of robustness.
In this section, we use some of those alternative measures to explain the robustness
of the proposed minimum LSD estimators.

\subsection{Higher Order Influence Analysis}

The higher (second) order influence function analysis for studying 
the robustness of a minimum divergence estimators
was used by Lindsay (1994) for the case of PD family and recently by Ghosh et al.~(2013) 
for the $S$-divergence family; both the work have shown this approach to provide significantly 
improved prediction of the robustness of corresponding estimators.  
Here, we present a similar analysis for the minimum LSD estimator.

For any functional $T$, $\Delta T(\epsilon) = T(G_\epsilon) - T(G)$
quantifies the amount of bias under contamination as a function of contamination proportion 
$\epsilon$, which can be approximated using the first-order Taylor expansion as 
$\Delta T(\epsilon) = T(G_\epsilon ) - T( G ) \approx \epsilon T'(y)$.
Hence the first order influence function gives an approximation to the predicted bias up to first order. 
When this first order approximation fails, we can consider a second order (approximate) bias prediction by 
$\Delta T(\epsilon) = \epsilon T'(y) + \frac{\epsilon^2}{2} T''(y)$.
The term $T''(y)$ is interpreted as the second order influence function and the ratio 
$$\frac{\mbox{quadratic approximation}}{\mbox{linear approximation}} 
= 1 + \frac{[T''(y)/T'(y)]\epsilon}{2}$$ 
serve as a  measure of adequacy of the first-order approximation and hence of the first order 
influence analysis; the two approximation may differ significantly 
for fairly small values of $\epsilon$ when the first order approximation is inadequate.
Our next theorem present the expression of the second order approximation $T''(y)$
for the minimum LSD estimator with a scalar parameter; this can be routinely extended to the case of 
vector parameter also. Let us define, for the model family $\{f_\theta\}$ with a scalar $\theta$,
the quantities   $c_i = \int u_\theta^i f_\theta^{1+\beta}$ and 
 $d_i = \int [\nabla u_\theta] u_\theta^i f_\theta^{1+\beta}$ for $i = 0, 1, 2, 3$.

\begin{theorem}
Under the above mentioned set-up with a scalar parameter $\theta$, 
if  true distribution belonging to the model family then
the second order influence function of the minimum LSD estimator defined by 
the estimating equation (\ref{EQ:lsd_est_eqn}) is  
$$T''(y) = (N_0^p D_0 - N_0 D_0^p)/D_0^2.$$
where  
\begin{eqnarray}
N_0 &=& f_\theta^\beta(y) u_\theta(y) c_o - f_\theta^\beta(y) c_1, \nonumber \\
D_0 &=& c_2 c_0 - c_1^2, \nonumber \\
N_0^p &=& B T'(y) c_1^2 -(A-1) c_1 c_0 + 2 (A-1) f_\theta^\beta(y) c_1 \nonumber \\
& & - B T'(y) f_\theta^\beta(y) u_\theta(y) c_1 -(A-1) f_\theta^{\beta-1}(y) c_1  \nonumber \\
& & - B T'(y) c_2 c_0 -T'(y)d_0 c_0 + B T'(y)f_\theta^\beta(y) u_\theta^2(y) c_0  \nonumber \\
& & + f_\theta^\beta(y) u_\theta'(y) T'(y) c_0 -(A-1) f_\theta^\beta(y) u_\theta(y) c_0 \nonumber \\
& & + (A-1) f_\theta^{\beta-1}(y) u_\theta(y) c_0 - (1+\beta) f_\theta^\beta(y) u_\theta(y) T'(y) c_1  \nonumber \\
& & + (1+\beta)  T'(y) c_0 c_2 - (1+\beta) T'(y) f_\theta^\beta(y) c_2  \nonumber \\
{\rm and}  \nonumber \\
D_0^p &=& (A^2+2AB) T'(y) c_3 c_0 + 3 A T'(y) d_1 c_0 + B(1+\beta) T'(y) c_2 c_0 -A^2 c_2 c_0  \nonumber \\
& & + A(1+\beta) f_\theta^\beta(y) c_2 -(A+\beta) T'(y) d_0 c_1 + A f_\theta^\beta(y) d_0 \nonumber \\
& & -(AB+A(1+\beta)+(1+\beta)B) T'(y) c_2 c_1 +A^2 c_1^2 -A^2 f_\theta^\beta(y) u_\theta(y) c_1  \nonumber \\
& & - AB f_\theta^\beta(y) u_\theta^2(y) c_0 -A f_\theta^\beta(y) u_\theta'(y)c_0, \nonumber 
\end{eqnarray}
where $u_\theta'(y)=\frac{\partial}{\partial \theta}u_\theta(y).$
\end{theorem}

\begin{figure}
\centering
\subfloat[$\beta=0$, $\gamma=0$]{
\includegraphics
[width=0.5\textwidth]
{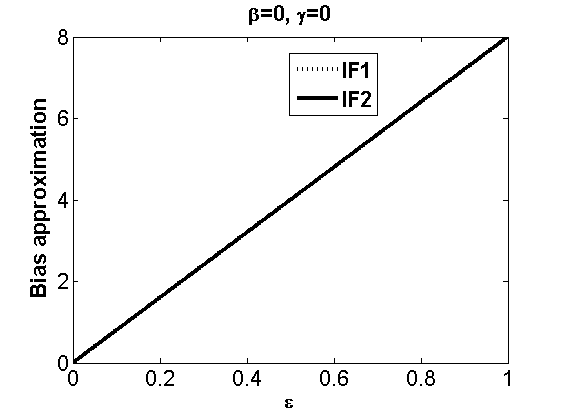}
\label{FIG:1a}}
~ 
\subfloat[$\beta=0.3$, $\gamma=0$]{
\includegraphics[width=0.5\textwidth]{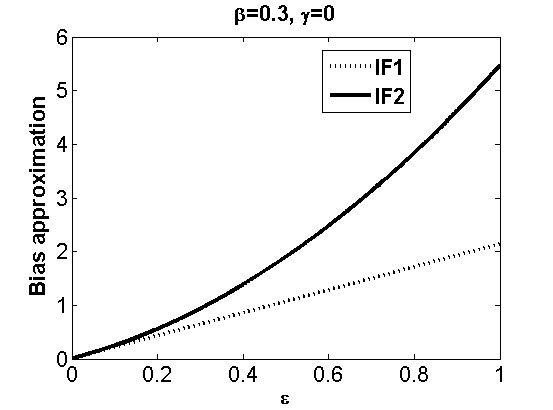}
\label{FIG:1b}}
\\
\subfloat[$\beta=0.6$, $\gamma=0$]{
\includegraphics[width=0.5\textwidth]{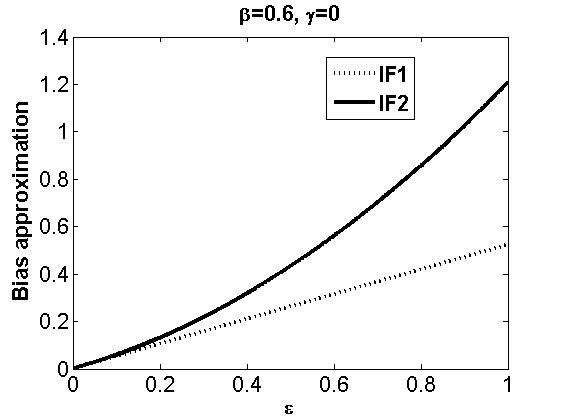}
\label{FIG:1c}}
~ 
\subfloat[$\beta=1$, $\gamma=0$]{
\includegraphics[width=0.5\textwidth]{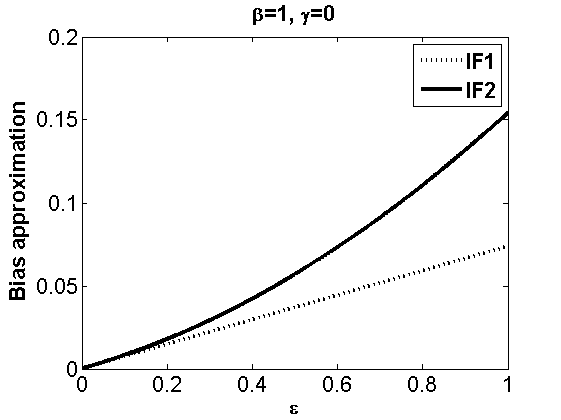}
\label{FIG:1d}}
\caption{Bias Approximations (solid line : second order; dashed line : first order) for the minimum LSD estimator
for $\gamma=0$}
\label{FIG:Second_Order_A}
\end{figure}
\begin{figure}
\centering
\subfloat[$\beta=0$, $\gamma=0.1$]{
\includegraphics[width=0.5\textwidth]{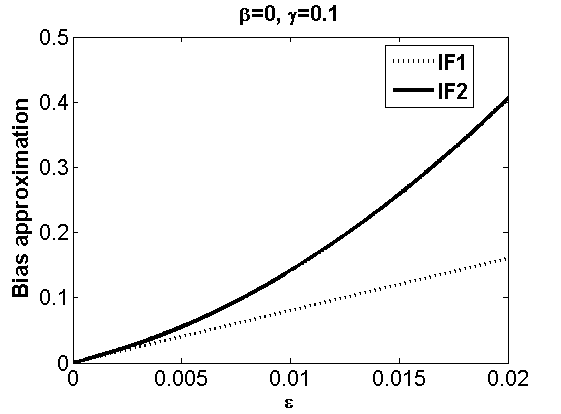}
\label{FIG:2a}}
~ 
\subfloat[$\beta=0.5$, $\gamma=0.1$]{
\includegraphics[width=0.5\textwidth]{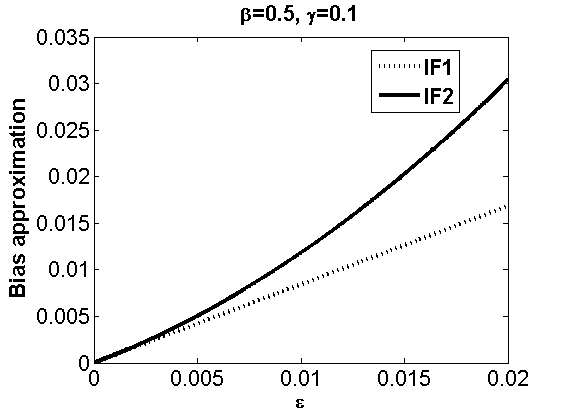}
\label{FIG:2b}}
\\
\subfloat[$\beta=0$, $\gamma=1$]{
\includegraphics[width=0.5\textwidth]{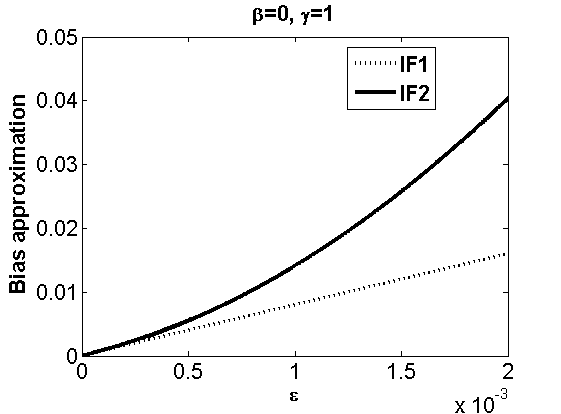}
\label{FIG:2c}}
~ 
\subfloat[$\beta=0.5$, $\gamma=1$]{
\includegraphics[width=0.5\textwidth]{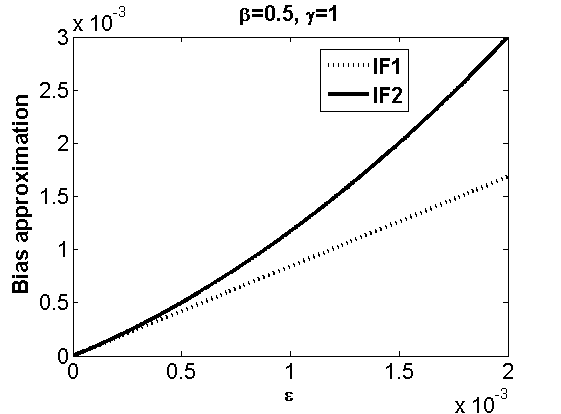}
\label{FIG:2d}}
\caption{Bias Approximations (solid line : second order; dashed line : first order) for the minimum LSD estimator
for $\gamma>0$}
\label{FIG:Second_Order_B}
\end{figure}
\begin{figure}
\centering
\subfloat[$\beta=0$, $\gamma=-0.1$]{
\includegraphics
[width=0.5\textwidth]
{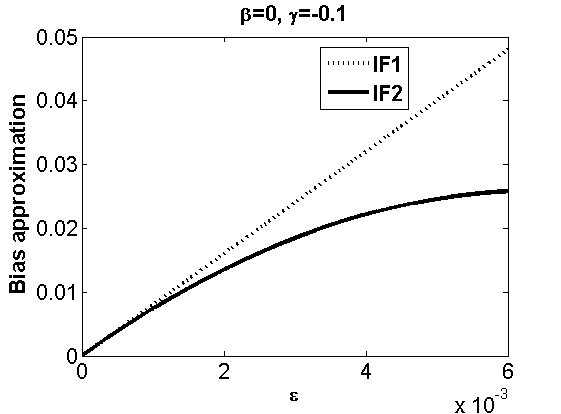}
\label{FIG:3a}}
~ 
\subfloat[$\beta=0.5$, $\gamma=-0.1$]{
\includegraphics[width=0.5\textwidth]{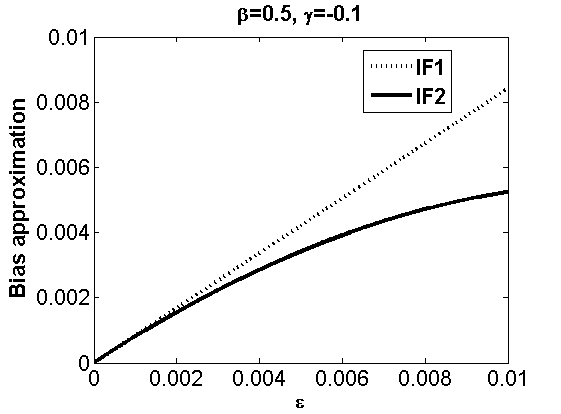}
\label{FIG:3b}}
\\
\subfloat[$\beta=0$, $\gamma=-1$]{
\includegraphics[width=0.5\textwidth]{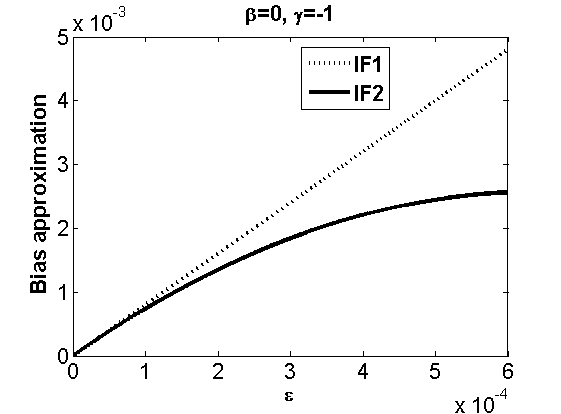}
\label{FIG:3c}}
~ 
\subfloat[$\beta=0.5$, $\gamma=-1$]{
\includegraphics[width=0.5\textwidth]{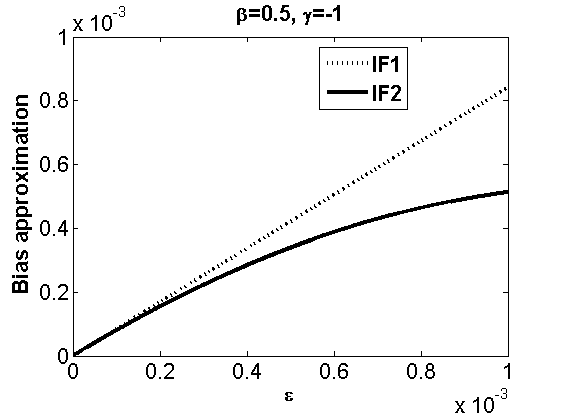}
\label{FIG:3d}}
\caption{Bias Approximations (solid line : second order; dashed line : first order) for the minimum LSD estimator
for $\gamma<0$}
\label{FIG:Second_Order_C}
\end{figure}

\noindent
\textbf{Example (Poisson Mean):}	Let us now consider a numerical simulation to study the 
performance of the above second order influence analysis through its application in case of 
the Poisson model with mean $\theta$. Using the special structure of one parameter exponential family,
of which Poisson distribution is a special case, we compute the first and second order 
bias approximation using their respective expressions as given above and in Section \ref{SEC:IF_MLSDE}.
However, for brevity, we will only present some particular simulation result 
with $\theta=4$, the contamination point $y=12$ and specific $(\beta,\gamma)$ combinations 
and the corresponding bias plots are shown in Figures \ref{FIG:Second_Order_A}, \ref{FIG:Second_Order_B} 
and \ref{FIG:Second_Order_C} respectively for $\gamma = 0$, $\gamma > 0 $ and $\gamma < 0$.

{\em Comments on Figure \ref{FIG:Second_Order_A} $(\gamma=0)$}: As 
expected both first order and second order influence function for $\beta=0$ 
gives a straight line. The bias approximation decreases as $\beta$ increases
for both first and second order influence function. The difference of approximation 
between first and second order decreases as $\beta$ increases. 

{\em Comments on Figure \ref{FIG:Second_Order_B} $(\gamma>0)$}: 
Keeping $\gamma$ fixed the difference between bias approximation 
among first and second order decreases as $\beta$ increases.   

{\em Comments on Figure \ref{FIG:Second_Order_C} $(\gamma<0)$}: 
As expected for this case the bias approximation is more for the first order influence function compared 
to the second order but the difference among two types of influence function 
shows same behavior compared to the case $\gamma>0$.  

\subsection{A Breakdown Point Result : Location Model}

Another popular alternative to the influence function analysis is the breakdown point theory;
following Simpson (1987) we will say that the estimator $T(G)$ breaks down 
for contamination level $\epsilon$ if $|T_{\beta,\gamma}(H_{\epsilon ,n}) - T (G)| \rightarrow \infty$  
as $n \rightarrow \infty$ for some sequence $K_n$ and $H_{\epsilon ,n} = (1 - \epsilon )G+\epsilon K_n$.
Although the derivation of a general breakdown result is difficult, 
several authors have used it for some suitable subclass of probability distributions; 
see Park and Basu (2004), Ghosh et al.~(2013) for breakdown results on some 
related minimum divergence estimators.

Now we derive the breakdown point of the minimum LSD functional \\ $T_{\beta, \gamma}(G)$ 
under the special class of location family 
$\mathcal{F}_\theta = \{ f_\theta(x) = f(x-\theta) : \theta \in \Theta \}$. 
The particular property of this family, that helps to make the calculations simpler, is 
$$\int \{f(x- \theta)\}^{1+\beta}dx = \int \{f(x)\}^{1+\beta}dx = M_f^\beta,~~~~\mbox{(say)}$$ 
which is independent of the parameter $\theta$. 
Using this and the increasing nature of the logarithmic function, 
the minimum LSD estimator for a location model is seen to be the maximizer of only the one integral term
$\int f^{B} g^{A}$ whenever $A>0$ and $B>0$. However, under the same location model the 
minimum $S$-divergence estimator of Ghosh et al.~(2013) can also be seen to be the maximizer of the same
integral. Therefore, under the location family of densities, the minimum LSD estimator 
with $A>0$ and $B>0$ coincides with corresponding minimum $S$-divergence estimators. 
Then it follows from Ghosh et al.~(2013) that,
under certain assumptions (assumptions BP1 to BP3 of their paper) the asymptotic 
breakdown point $\epsilon^*$ of the minimum LSD estimator  $T_{\beta,\gamma}$ 
with $A>0$ and $B>0$ is at least $\frac{1}{2}$ at the model family.
\section{Testing Hypotheses Simulation}
\label{SEC:testing_simulation}
This section will describe the testing of hypotheses simulation example. 
We have taken sample from 
$(1-\epsilon) \mbox{ Poisson}(2)+\epsilon \mbox{ Poisson}(15)$ for $\epsilon=0,0.1$
and various sample sizes $n=20,50,100$. All simulations have been replicated
$1000$ times. Tables \ref{TAB:level_20_nc},
\ref{TAB:level_50_nc} and \ref{TAB:level_100_nc} give us the observed levels
for no contamination case and tables 
\ref{TAB:level_20},
\ref{TAB:level_50} and \ref{TAB:level_100} 
for contamination case
while testing 
$H_0:\theta=2$ and the powers given in tables
\ref{TAB:power_20_nc},
\ref{TAB:power_50_nc} and \ref{TAB:power_100_nc}
for no contamination case and 
\ref{TAB:power_20},
\ref{TAB:power_50} and \ref{TAB:power_100}
for contamination case
considering the testing problem
$H_0:\theta=3$. Here the observed level has been taken as $0.05$. 
Usually for both $\beta$ and $\gamma$ close to $0$ we get 
level close to $0.05$ under no contamination case. For $\beta \geq 0.04$, 
level does not go under $0.06$ for any $\gamma$. As $\gamma$ becomes 
distant from $0$ in both positive and negative direction level 
moves from $0.05$ under no contamination. Under contamination 
set-up, empirical level usually does not go below than $0.1$. 
For smaller sample size like $20$ level never become lower than $0.1$
whereas for large sample size as $100$, only when $-0.5\leq\gamma\leq-0.3$
and $\beta\leq0.2$ level becomes lower than $0.1$ and for moderately 
large sample size $50$, the situation does not differ very significantly. 
Under contamination set-up for $\beta\leq0.4$ and $\gamma>0$, the level 
is very high and for sample size $100$ it goes to $1$ also. The empirical 
power is very high under no contamination. For sample size $100$ the 
power is $1$ for most of the values of $\beta$ and $\gamma$. Though for 
sample size $50$ the power does not reach $1$ but it is usually very close to $1$.
Power usually does not go to that close to $1$ for sample size $20$ except for 
high negative value of $\gamma (\leq-0.5)$ and lower value of $\beta (\leq 0.2)$. 
Under contamination set-up and for sample size $20$ the power usually 
does not go to $1$ but for low $\beta$ and high negative value of $\gamma$ it 
goes very close. For $\gamma$ close to $0$ and low $\beta$, the power 
becomes less than $0.5$ but this is not much common throughout the table. 
For sample size $100$, the power is usually becomes $1$ except for very few 
combinations of $(\beta,\gamma)$ and this fact is maintained for sample size 
$50$ also. 
As shown earlier, for $\beta=1$, the divergence is independent of $\gamma$,
that fact is also evident from the result that both level and power 
for all values of $\gamma$ is same for $\beta=1$. 
\begin{table}
\caption{\label{TAB:level_20_nc}
The Empirical level of the LSD statistic under $H_0:\theta=2$ for different values of $\beta$ and $\gamma$ ($n=20$)
(No Contamination)}
\centering
\resizebox{\textwidth}{!}{ %
\begin{tabular}{r r r r r r r r r} \hline
$\gamma$	&	$\beta=0$	&	$\beta=0.1$	&	$\beta=0.2$	&	$\beta=0.4$	&	$\beta=0.7$	&	$\beta=0.8$	&	$\beta=0.9$	&	$\beta=1$	\\ \hline 
$-1$ & -- & 0.629 & 0.35 & 0.171 & 0.119 & 0.123 & 0.131 & 0.135 \\ 
$-0.9$ & 0.673 & 0.418 & 0.264 & 0.143 & 0.118 & 0.123 & 0.13 & 0.135 \\ 
$-0.7$ & 0.294 & 0.209 & 0.165 & 0.126 & 0.115 & 0.123 & 0.131 & 0.135 \\ 
$-0.5$ & 0.149 & 0.133 & 0.118 & 0.111 & 0.113 & 0.123 & 0.13 & 0.135 \\ 
$-0.3$ & 0.104 & 0.098 & 0.101 & 0.103 & 0.111 & 0.121 & 0.128 & 0.135 \\ 
$-0.1$ & 0.078 & 0.084 & 0.083 & 0.097 & 0.109 & 0.12 & 0.127 & 0.135 \\ 
0 & 0.041 & 0.08 & 0.081 & 0.093 & 0.11 & 0.118 & 0.127 & 0.135 \\ 
0.1 & 0.073 & 0.078 & 0.079 & 0.092 & 0.112 & 0.118 & 0.126 & 0.135 \\ 
0.3 & 0.074 & 0.072 & 0.074 & 0.085 & 0.113 & 0.115 & 0.126 & 0.135 \\ 
0.5 & 0.081 & 0.076 & 0.075 & 0.087 & 0.109 & 0.117 & 0.127 & 0.135 \\ 
0.7 & 0.101 & 0.085 & 0.078 & 0.083 & 0.109 & 0.117 & 0.128 & 0.135 \\ 
0.9 & 0.118 & 0.1 & 0.084 & 0.083 & 0.108 & 0.119 & 0.128 & 0.135 \\ 
1 & 0.128 & 0.111 & 0.089 & 0.083 & 0.109 & 0.119 & 0.127 & 0.135 \\ 
1.5 & 0.159 & 0.154 & 0.13 & 0.089 & 0.105 & 0.114 & 0.127 & 0.135 \\ 
2 & 0.16 & 0.164 & 0.163 & 0.112 & 0.103 & 0.112 & 0.122 & 0.135 \\ 

\hline
\end{tabular}}
\end{table}

\begin{table}
\caption{\label{TAB:level_50_nc}
The Empirical level of the LSD statistic under $H_0:\theta=2$ for different values of $\beta$ and $\gamma$ ($n=50$)
(No Contamination)}
\centering
\resizebox{\textwidth}{!}{ %
\begin{tabular}{r r r r r r r r r} \hline
$\gamma$	&	$\beta=0$	&	$\beta=0.1$	&	$\beta=0.2$	&	$\beta=0.4$	&	$\beta=0.7$	&	$\beta=0.8$	&	$\beta=0.9$	&	$\beta=1$	\\ \hline 

$-1$ & -- & 0.439 & 0.204 & 0.093 & 0.079 & 0.091 & 0.099 & 0.102 \\ 
$-0.9$ & 0.56 & 0.241 & 0.142 & 0.086 & 0.078 & 0.089 & 0.097 & 0.102 \\ 
$-0.7$ & 0.187 & 0.118 & 0.089 & 0.074 & 0.077 & 0.087 & 0.097 & 0.102 \\ 
$-0.5$ & 0.09 & 0.078 & 0.07 & 0.064 & 0.077 & 0.087 & 0.097 & 0.102 \\ 
$-0.3$ & 0.066 & 0.063 & 0.063 & 0.061 & 0.078 & 0.085 & 0.097 & 0.102 \\ 
$-0.1$ & 0.053 & 0.054 & 0.059 & 0.063 & 0.077 & 0.084 & 0.096 & 0.102 \\ 
0 & 0.05 & 0.053 & 0.057 & 0.061 & 0.076 & 0.084 & 0.096 & 0.102 \\ 
0.1 & 0.047 & 0.054 & 0.055 & 0.06 & 0.075 & 0.085 & 0.096 & 0.102 \\ 
0.3 & 0.054 & 0.053 & 0.058 & 0.061 & 0.076 & 0.084 & 0.095 & 0.102 \\ 
0.5 & 0.068 & 0.059 & 0.057 & 0.062 & 0.074 & 0.084 & 0.094 & 0.102 \\ 
0.7 & 0.093 & 0.073 & 0.061 & 0.061 & 0.073 & 0.084 & 0.094 & 0.102 \\ 
0.9 & 0.12 & 0.09 & 0.07 & 0.06 & 0.073 & 0.083 & 0.093 & 0.102 \\ 
1 & 0.123 & 0.102 & 0.075 & 0.061 & 0.074 & 0.083 & 0.093 & 0.102 \\ 
1.5 & 0.157 & 0.139 & 0.117 & 0.068 & 0.073 & 0.084 & 0.091 & 0.102 \\ 
2 & 0.218 & 0.165 & 0.145 & 0.084 & 0.073 & 0.08 & 0.092 & 0.102 \\ 

\hline
\end{tabular}}
\end{table}

\begin{table}
\caption{\label{TAB:level_100_nc}
The Empirical level of the LSD statistic under $H_0:\theta=2$ for different values of $\beta$ and $\gamma$ ($n=100$)
(No Contamination)}
\centering
\resizebox{\textwidth}{!}{ %
\begin{tabular}{r r r r r r r r r} \hline
$\gamma$	&	$\beta=0$	&	$\beta=0.1$	&	$\beta=0.2$	&	$\beta=0.4$	&	$\beta=0.7$	&	$\beta=0.8$	&	$\beta=0.9$	&	$\beta=1$	\\ \hline 

$-1$ & -- & 0.373 & 0.141 & 0.087 & 0.102 & 0.111 & 0.117 & 0.127 \\ 
$-0.9$ & 0.444 & 0.176 & 0.114 & 0.08 & 0.102 & 0.111 & 0.117 & 0.127 \\ 
$-0.7$ & 0.131 & 0.098 & 0.083 & 0.078 & 0.1 & 0.11 & 0.118 & 0.127 \\ 
$-0.5$ & 0.078 & 0.071 & 0.076 & 0.075 & 0.099 & 0.108 & 0.119 & 0.127 \\ 
$-0.3$ & 0.061 & 0.061 & 0.066 & 0.079 & 0.099 & 0.108 & 0.119 & 0.127 \\ 
$-0.1$ & 0.054 & 0.057 & 0.063 & 0.076 & 0.099 & 0.109 & 0.12 & 0.127 \\ 
0 & 0.048 & 0.057 & 0.06 & 0.075 & 0.1 & 0.109 & 0.12 & 0.127 \\ 
0.1 & 0.054 & 0.055 & 0.06 & 0.076 & 0.1 & 0.109 & 0.12 & 0.127 \\ 
0.3 & 0.062 & 0.06 & 0.061 & 0.077 & 0.099 & 0.109 & 0.12 & 0.127 \\ 
0.5 & 0.075 & 0.065 & 0.058 & 0.077 & 0.1 & 0.109 & 0.118 & 0.127 \\ 
0.7 & 0.082 & 0.075 & 0.062 & 0.078 & 0.099 & 0.109 & 0.119 & 0.127 \\ 
0.9 & 0.112 & 0.08 & 0.072 & 0.077 & 0.098 & 0.11 & 0.118 & 0.127 \\ 
1 & 0.122 & 0.09 & 0.077 & 0.078 & 0.098 & 0.11 & 0.118 & 0.127 \\ 
1.5 & 0.196 & 0.146 & 0.108 & 0.08 & 0.095 & 0.109 & 0.116 & 0.127 \\ 
2 & 0.231 & 0.208 & 0.155 & 0.095 & 0.096 & 0.107 & 0.116 & 0.127 \\

\hline
\end{tabular}}
\end{table}
\begin{table}
\caption{\label{TAB:power_20_nc}
The Empirical power of the LSD statistic under $H_0:\theta=3$ 
against $H_1:\theta=2$ for different values of $\beta$ and $\gamma$ ($n=20$)(No Contamination)}
\centering
\resizebox{\textwidth}{!}{ %
\begin{tabular}{r r r r r r r r r} \hline
$\gamma$	&	$\beta=0$	&	$\beta=0.1$	&	$\beta=0.2$	&	$\beta=0.4$	&	$\beta=0.7$	&	$\beta=0.8$	&	$\beta=0.9$	&	$\beta=1$	\\ \hline 

$-1$ & -- & 0.997 & 0.98 & 0.948 & 0.885 & 0.865 & 0.853 & 0.834 \\ 
$-0.9$ & 0.998 & 0.991 & 0.969 & 0.939 & 0.88 & 0.864 & 0.851 & 0.834 \\ 
$-0.7$ & 0.973 & 0.965 & 0.956 & 0.917 & 0.874 & 0.861 & 0.85 & 0.834 \\ 
$-0.5$ & 0.955 & 0.938 & 0.922 & 0.895 & 0.865 & 0.856 & 0.848 & 0.834 \\ 
$-0.3$ & 0.912 & 0.899 & 0.889 & 0.876 & 0.858 & 0.851 & 0.847 & 0.834 \\ 
$-0.1$ & 0.868 & 0.865 & 0.863 & 0.86 & 0.854 & 0.85 & 0.846 & 0.834 \\ 
0 & 0.828 & 0.85 & 0.853 & 0.855 & 0.853 & 0.85 & 0.846 & 0.834 \\ 
0.1 & 0.838 & 0.838 & 0.846 & 0.847 & 0.851 & 0.849 & 0.846 & 0.834 \\ 
0.3 & 0.808 & 0.816 & 0.826 & 0.84 & 0.847 & 0.85 & 0.843 & 0.834 \\ 
0.5 & 0.766 & 0.791 & 0.805 & 0.827 & 0.841 & 0.842 & 0.842 & 0.834 \\ 
0.7 & 0.739 & 0.759 & 0.778 & 0.808 & 0.829 & 0.836 & 0.84 & 0.834 \\ 
0.9 & 0.721 & 0.736 & 0.753 & 0.797 & 0.827 & 0.833 & 0.838 & 0.834 \\ 
1 & 0.704 & 0.725 & 0.746 & 0.791 & 0.826 & 0.831 & 0.837 & 0.834 \\ 
1.5 & 0.647 & 0.677 & 0.701 & 0.754 & 0.813 & 0.824 & 0.831 & 0.834 \\ 
2 & 0.609 & 0.637 & 0.67 & 0.726 & 0.795 & 0.814 & 0.825 & 0.834 \\

\hline
\end{tabular}}
\end{table}
\begin{table}
\caption{\label{TAB:power_50_nc}
The Empirical power of the LSD statistic under $H_0:\theta=3$ 
against $H_1:\theta=2$ for different values of $\beta$ and $\gamma$ ($n=50$)(No Contamination)}
\centering
\resizebox{\textwidth}{!}{ %
\begin{tabular}{r r r r r r r r r} \hline
$\gamma$	&	$\beta=0$	&	$\beta=0.1$	&	$\beta=0.2$	&	$\beta=0.4$	&	$\beta=0.7$	&	$\beta=0.8$	&	$\beta=0.9$	&	$\beta=1$	\\ \hline 

$-1$ & -- & 0.999 & 0.999 & 0.997 & 0.994 & 0.991 & 0.99 & 0.987 \\ 
$-0.9$ & 0.999 & 0.999 & 0.998 & 0.996 & 0.994 & 0.991 & 0.99 & 0.987 \\ 
$-0.7$ & 0.999 & 0.998 & 0.998 & 0.995 & 0.994 & 0.991 & 0.99 & 0.987 \\ 
$-0.5$ & 0.997 & 0.996 & 0.997 & 0.995 & 0.992 & 0.991 & 0.99 & 0.987 \\ 
$-0.3$ & 0.996 & 0.996 & 0.995 & 0.994 & 0.991 & 0.991 & 0.99 & 0.987 \\ 
$-0.1$ & 0.996 & 0.994 & 0.994 & 0.993 & 0.991 & 0.99 & 0.989 & 0.987 \\ 
0 & 0.994 & 0.995 & 0.993 & 0.993 & 0.991 & 0.99 & 0.989 & 0.987 \\ 
0.1 & 0.995 & 0.994 & 0.992 & 0.993 & 0.991 & 0.99 & 0.989 & 0.987 \\ 
0.3 & 0.992 & 0.994 & 0.993 & 0.993 & 0.991 & 0.99 & 0.989 & 0.987 \\ 
0.5 & 0.99 & 0.991 & 0.993 & 0.992 & 0.99 & 0.989 & 0.989 & 0.987 \\ 
0.7 & 0.986 & 0.989 & 0.99 & 0.991 & 0.989 & 0.989 & 0.989 & 0.987 \\ 
0.9 & 0.977 & 0.988 & 0.987 & 0.99 & 0.989 & 0.989 & 0.989 & 0.987 \\ 
1 & 0.975 & 0.986 & 0.987 & 0.989 & 0.989 & 0.989 & 0.989 & 0.987 \\ 
1.5 & 0.963 & 0.971 & 0.981 & 0.988 & 0.989 & 0.988 & 0.988 & 0.987 \\ 
2 & 0.951 & 0.96 & 0.967 & 0.985 & 0.988 & 0.988 & 0.988 & 0.987 \\

\hline
\end{tabular}}
\end{table}
\begin{table}
\caption{\label{TAB:power_100_nc}
The Empirical power of the LSD statistic under $H_0:\theta=3$ 
against $H_1:\theta=2$ for different values of $\beta$ and $\gamma$ ($n=100$)(No Contamination)}
\centering
\resizebox{\textwidth}{!}{ %
\begin{tabular}{r r r r r r r r r} \hline
$\gamma$	&	$\beta=0$	&	$\beta=0.1$	&	$\beta=0.2$	&	$\beta=0.4$	&	$\beta=0.7$	&	$\beta=0.8$	&	$\beta=0.9$	&	$\beta=1$	\\ \hline 

$-1$ & -- & 1 & 1 & 1 & 1 & 1 & 1 & 1 \\ 
$-0.9$ & 1 & 1 & 1 & 1 & 1 & 1 & 1 & 1 \\ 
$-0.7$ & 1 & 1 & 1 & 1 & 1 & 1 & 1 & 1 \\ 
$-0.5$ & 1 & 1 & 1 & 1 & 1 & 1 & 1 & 1 \\ 
$-0.3$ & 1 & 1 & 1 & 1 & 1 & 1 & 1 & 1 \\ 
$-0.1$ & 1 & 1 & 1 & 1 & 1 & 1 & 1 & 1 \\ 
0 & 1 & 1 & 1 & 1 & 1 & 1 & 1 & 1 \\ 
0.1 & 1 & 1 & 1 & 1 & 1 & 1 & 1 & 1 \\ 
0.3 & 1 & 1 & 1 & 1 & 1 & 1 & 1 & 1 \\ 
0.5 & 1 & 1 & 1 & 1 & 1 & 1 & 1 & 1 \\ 
0.7 & 0.999 & 1 & 1 & 1 & 1 & 1 & 1 & 1 \\ 
0.9 & 0.998 & 0.999 & 1 & 1 & 1 & 1 & 1 & 1 \\ 
1 & 0.995 & 0.999 & 1 & 1 & 1 & 1 & 1 & 1 \\ 
1.5 & 0.993 & 0.994 & 0.999 & 1 & 1 & 1 & 1 & 1 \\ 
2 & 0.987 & 0.993 & 0.994 & 1 & 1 & 1 & 1 & 1 \\ 

\hline
\end{tabular}}
\end{table}
\begin{table}
\caption{\label{TAB:level_20}
The Empirical level of the LSD statistic under $H_0:\theta=2$ for different values of $\beta$ and $\gamma$ ($n=20$)
($90\%{\rm Poisson}(2)+10\%{\rm Poisson}(15)$)}
\centering
\resizebox{\textwidth}{!}{ %
\begin{tabular}{r r r r r r r r r} \hline
$\gamma$	&	$\beta=0$	&	$\beta=0.1$	&	$\beta=0.2$	&	$\beta=0.4$	&	$\beta=0.7$	&	$\beta=0.8$	&	$\beta=0.9$	&	$\beta=1$	\\ \hline 
$-1$ & -- & 0.647 & 0.388 & 0.205 & 0.162 & 0.164 & 0.171 & 0.173 \\ 
$-0.9$ & 0.685 & 0.435 & 0.289 & 0.177 & 0.16 & 0.164 & 0.171 & 0.173 \\ 
$-0.7$ & 0.316 & 0.242 & 0.194 & 0.156 & 0.153 & 0.16 & 0.168 & 0.173 \\ 
$-0.5$ & 0.175 & 0.161 & 0.15 & 0.142 & 0.151 & 0.159 & 0.167 & 0.173 \\ 
$-0.3$ & 0.122 & 0.126 & 0.126 & 0.133 & 0.15 & 0.157 & 0.164 & 0.173 \\ 
$-0.1$ & 0.233 & 0.112 & 0.113 & 0.13 & 0.145 & 0.156 & 0.163 & 0.173 \\ 
0 & 0.684 & 0.216 & 0.119 & 0.126 & 0.144 & 0.152 & 0.162 & 0.173 \\ 
0.1 & 0.863 & 0.637 & 0.15 & 0.127 & 0.141 & 0.152 & 0.161 & 0.173 \\ 
0.3 & 0.883 & 0.87 & 0.778 & 0.123 & 0.14 & 0.152 & 0.159 & 0.173 \\ 
0.5 & 0.885 & 0.881 & 0.87 & 0.178 & 0.14 & 0.148 & 0.157 & 0.173 \\ 
0.7 & 0.889 & 0.884 & 0.88 & 0.653 & 0.14 & 0.146 & 0.157 & 0.173 \\ 
0.9 & 0.892 & 0.887 & 0.883 & 0.849 & 0.141 & 0.145 & 0.156 & 0.173 \\ 
1 & 0.894 & 0.891 & 0.885 & 0.865 & 0.141 & 0.145 & 0.156 & 0.173 \\ 
1.5 & 0.897 & 0.896 & 0.892 & 0.882 & 0.138 & 0.146 & 0.154 & 0.173 \\ 
2 & 0.897 & 0.898 & 0.897 & 0.888 & 0.17 & 0.147 & 0.155 & 0.173 \\

\hline
\end{tabular}}
\end{table}

\begin{table}
\caption{\label{TAB:level_50}
The Empirical level of the LSD statistic under $H_0:\theta=2$ for different values of $\beta$ and $\gamma$ ($n=50$)
($90\%{\rm Poisson}(2)+10\%{\rm Poisson}(15)$)}
\centering
\resizebox{\textwidth}{!}{ %
\begin{tabular}{r r r r r r r r r} \hline
$\gamma$	&	$\beta=0$	&	$\beta=0.1$	&	$\beta=0.2$	&	$\beta=0.4$	&	$\beta=0.7$	&	$\beta=0.8$	&	$\beta=0.9$	&	$\beta=1$	\\ \hline 

$-1$ & -- & 0.436 & 0.214 & 0.111 & 0.105 & 0.114 & 0.128 & 0.14 \\ 
$-0.9$ & 0.545 & 0.256 & 0.16 & 0.102 & 0.104 & 0.114 & 0.128 & 0.14 \\ 
$-0.7$ & 0.191 & 0.125 & 0.106 & 0.085 & 0.103 & 0.113 & 0.128 & 0.14 \\ 
$-0.5$ & 0.099 & 0.094 & 0.088 & 0.083 & 0.102 & 0.112 & 0.128 & 0.14 \\ 
$-0.3$ & 0.087 & 0.078 & 0.078 & 0.082 & 0.103 & 0.111 & 0.126 & 0.14 \\ 
$-0.1$ & 0.402 & 0.123 & 0.08 & 0.081 & 0.102 & 0.109 & 0.126 & 0.14 \\ 
0 & 0.937 & 0.333 & 0.102 & 0.083 & 0.101 & 0.11 & 0.125 & 0.14 \\ 
0.1 & 0.986 & 0.867 & 0.222 & 0.085 & 0.101 & 0.109 & 0.125 & 0.14 \\ 
0.3 & 0.995 & 0.987 & 0.951 & 0.093 & 0.103 & 0.108 & 0.125 & 0.14 \\ 
0.5 & 0.996 & 0.995 & 0.988 & 0.213 & 0.101 & 0.108 & 0.124 & 0.14 \\ 
0.7 & 0.996 & 0.996 & 0.994 & 0.824 & 0.099 & 0.106 & 0.122 & 0.14 \\ 
0.9 & 0.996 & 0.996 & 0.996 & 0.979 & 0.1 & 0.106 & 0.122 & 0.14 \\ 
1 & 0.996 & 0.996 & 0.996 & 0.986 & 0.1 & 0.106 & 0.12 & 0.14 \\ 
1.5 & 0.998 & 0.998 & 0.997 & 0.996 & 0.103 & 0.105 & 0.12 & 0.14 \\ 
2 & 0.998 & 0.998 & 0.998 & 0.997 & 0.139 & 0.104 & 0.119 & 0.14 \\ 

\hline
\end{tabular}}
\end{table}

\begin{table}
\caption{\label{TAB:level_100}
The Empirical level of the LSD statistic under $H_0:\theta=2$ for different values of $\beta$ and $\gamma$ ($n=100$)
($90\%{\rm Poisson}(2)+10\%{\rm Poisson}(15)$)}
\centering
\resizebox{\textwidth}{!}{ %
\begin{tabular}{r r r r r r r r r} \hline
$\gamma$	&	$\beta=0$	&	$\beta=0.1$	&	$\beta=0.2$	&	$\beta=0.4$	&	$\beta=0.7$	&	$\beta=0.8$	&	$\beta=0.9$	&	$\beta=1$	\\ \hline 

$-1$ & -- & 0.371 & 0.174 & 0.107 & 0.118 & 0.128 & 0.138 & 0.141 \\ 
$-0.9$ & 0.434 & 0.202 & 0.131 & 0.111 & 0.117 & 0.128 & 0.137 & 0.141 \\ 
$-0.7$ & 0.152 & 0.108 & 0.094 & 0.112 & 0.116 & 0.129 & 0.136 & 0.141 \\ 
$-0.5$ & 0.09 & 0.086 & 0.092 & 0.106 & 0.117 & 0.129 & 0.136 & 0.141 \\ 
$-0.3$ & 0.095 & 0.091 & 0.087 & 0.101 & 0.118 & 0.128 & 0.136 & 0.141 \\ 
$-0.1$ & 0.625 & 0.167 & 0.103 & 0.101 & 0.117 & 0.128 & 0.136 & 0.141 \\ 
0 & 0.996 & 0.488 & 0.131 & 0.099 & 0.117 & 0.126 & 0.136 & 0.141 \\ 
0.1 & 1 & 0.969 & 0.293 & 0.103 & 0.116 & 0.125 & 0.136 & 0.141 \\ 
0.3 & 1 & 1 & 0.997 & 0.112 & 0.116 & 0.125 & 0.135 & 0.141 \\ 
0.5 & 1 & 1 & 1 & 0.236 & 0.115 & 0.126 & 0.134 & 0.141 \\ 
0.7 & 1 & 1 & 1 & 0.914 & 0.115 & 0.125 & 0.135 & 0.141 \\ 
0.9 & 1 & 1 & 1 & 0.999 & 0.115 & 0.124 & 0.135 & 0.141 \\ 
1 & 1 & 1 & 1 & 0.999 & 0.116 & 0.124 & 0.135 & 0.141 \\ 
1.5 & 1 & 1 & 1 & 1 & 0.118 & 0.124 & 0.134 & 0.141 \\ 
2 & 1 & 1 & 1 & 1 & 0.164 & 0.123 & 0.133 & 0.141 \\

\hline
\end{tabular}}
\end{table}
\begin{table}
\caption{\label{TAB:power_20}
The Empirical power of the LSD statistic under $H_0:\theta=3$ 
against $H_1:\theta=2$ for different values of $\beta$ and $\gamma$ ($n=20$)
($90\%{\rm Poisson}(2)+10\%{\rm Poisson}(15)$)}
\centering
\resizebox{\textwidth}{!}{ %
\begin{tabular}{r r r r r r r r r} \hline
$\gamma$	&	$\beta=0$	&	$\beta=0.1$	&	$\beta=0.2$	&	$\beta=0.4$	&	$\beta=0.7$	&	$\beta=0.8$	&	$\beta=0.9$	&	$\beta=1$	\\ \hline 

$-1$ & -- & 1 & 0.983 & 0.943 & 0.848 & 0.829 & 0.807 & 0.797 \\ 
$-0.9$ & 0.999 & 0.99 & 0.973 & 0.931 & 0.845 & 0.828 & 0.807 & 0.797 \\ 
$-0.7$ & 0.982 & 0.964 & 0.942 & 0.897 & 0.836 & 0.824 & 0.806 & 0.797 \\ 
$-0.5$ & 0.935 & 0.92 & 0.909 & 0.868 & 0.828 & 0.816 & 0.805 & 0.797 \\ 
$-0.3$ & 0.841 & 0.853 & 0.853 & 0.844 & 0.821 & 0.81 & 0.804 & 0.797 \\ 
$-0.1$ & 0.471 & 0.701 & 0.786 & 0.816 & 0.809 & 0.805 & 0.801 & 0.797 \\ 
0 & 0.411 & 0.478 & 0.701 & 0.803 & 0.81 & 0.801 & 0.8 & 0.797 \\ 
0.1 & 0.773 & 0.34 & 0.549 & 0.786 & 0.806 & 0.8 & 0.798 & 0.797 \\ 
0.3 & 0.898 & 0.837 & 0.511 & 0.722 & 0.799 & 0.798 & 0.797 & 0.797 \\ 
0.5 & 0.925 & 0.902 & 0.84 & 0.54 & 0.791 & 0.793 & 0.798 & 0.797 \\ 
0.7 & 0.933 & 0.926 & 0.899 & 0.361 & 0.782 & 0.791 & 0.795 & 0.797 \\ 
0.9 & 0.932 & 0.932 & 0.918 & 0.729 & 0.774 & 0.789 & 0.795 & 0.797 \\ 
1 & 0.934 & 0.935 & 0.926 & 0.807 & 0.766 & 0.785 & 0.794 & 0.797 \\ 
1.5 & 0.937 & 0.934 & 0.934 & 0.906 & 0.728 & 0.772 & 0.785 & 0.797 \\ 
2 & 0.932 & 0.937 & 0.933 & 0.925 & 0.594 & 0.76 & 0.78 & 0.797 \\

\hline
\end{tabular}}
\end{table}
\begin{table}
\caption{\label{TAB:power_50}
The Empirical power of the LSD statistic under $H_0:\theta=3$ 
against $H_1:\theta=2$ for different values of $\beta$ and $\gamma$ ($n=50$)
($90\%{\rm Poisson}(2)+10\%{\rm Poisson}(15)$)}
\centering
\resizebox{\textwidth}{!}{ %
\begin{tabular}{r r r r r r r r r} \hline
$\gamma$	&	$\beta=0$	&	$\beta=0.1$	&	$\beta=0.2$	&	$\beta=0.4$	&	$\beta=0.7$	&	$\beta=0.8$	&	$\beta=0.9$	&	$\beta=1$	\\ \hline 

$-1$ & -- & 1 & 1 & 0.997 & 0.989 & 0.984 & 0.98 & 0.98 \\ 
$-0.9$ & 1 & 0.999 & 1 & 0.995 & 0.989 & 0.984 & 0.98 & 0.98 \\ 
$-0.7$ & 0.999 & 0.998 & 0.996 & 0.994 & 0.987 & 0.983 & 0.98 & 0.98 \\ 
$-0.5$ & 0.995 & 0.995 & 0.994 & 0.991 & 0.986 & 0.984 & 0.98 & 0.98 \\ 
$-0.3$ & 0.979 & 0.989 & 0.991 & 0.99 & 0.985 & 0.983 & 0.98 & 0.98 \\ 
$-0.1$ & 0.639 & 0.936 & 0.977 & 0.987 & 0.984 & 0.983 & 0.98 & 0.98 \\ 
0 & 0.452 & 0.712 & 0.953 & 0.985 & 0.984 & 0.983 & 0.98 & 0.98 \\ 
0.1 & 0.892 & 0.354 & 0.854 & 0.98 & 0.984 & 0.983 & 0.98 & 0.98 \\ 
0.3 & 0.982 & 0.944 & 0.501 & 0.973 & 0.984 & 0.983 & 0.98 & 0.98 \\ 
0.5 & 0.991 & 0.983 & 0.945 & 0.89 & 0.982 & 0.983 & 0.98 & 0.98 \\ 
0.7 & 0.992 & 0.99 & 0.984 & 0.386 & 0.981 & 0.982 & 0.98 & 0.98 \\ 
0.9 & 0.995 & 0.993 & 0.987 & 0.799 & 0.981 & 0.982 & 0.98 & 0.98 \\ 
1 & 0.995 & 0.992 & 0.992 & 0.904 & 0.981 & 0.982 & 0.979 & 0.98 \\ 
1.5 & 0.997 & 0.995 & 0.996 & 0.985 & 0.978 & 0.98 & 0.98 & 0.98 \\ 
2 & 0.998 & 0.998 & 0.996 & 0.994 & 0.94 & 0.98 & 0.98 & 0.98 \\

\hline
\end{tabular}}
\end{table}
\begin{table}
\caption{\label{TAB:power_100}
The Empirical power of the LSD statistic under $H_0:\theta=3$ 
against $H_1:\theta=2$ for different values of $\beta$ and $\gamma$ ($n=100$)
($90\%{\rm Poisson}(2)+10\%{\rm Poisson}(15)$)}
\centering
\resizebox{\textwidth}{!}{ %
\begin{tabular}{r r r r r r r r r} \hline
$\gamma$	&	$\beta=0$	&	$\beta=0.1$	&	$\beta=0.2$	&	$\beta=0.4$	&	$\beta=0.7$	&	$\beta=0.8$	&	$\beta=0.9$	&	$\beta=1$	\\ \hline 

$-1$ & -- & 1 & 1 & 1 & 1 & 1 & 1 & 1 \\ 
$-0.9$ & 1 & 1 & 1 & 1 & 1 & 1 & 1 & 1 \\ 
$-0.7$ & 1 & 1 & 1 & 1 & 1 & 1 & 1 & 1 \\ 
$-0.5$ & 1 & 1 & 1 & 1 & 1 & 1 & 1 & 1 \\ 
$-0.3$ & 0.998 & 0.999 & 1 & 1 & 1 & 1 & 1 & 1 \\ 
$-0.1$ & 0.818 & 0.996 & 0.999 & 1 & 1 & 1 & 1 & 1 \\ 
0 & 0.489 & 0.92 & 0.997 & 1 & 1 & 1 & 1 & 1 \\ 
0.1 & 0.97 & 0.358 & 0.983 & 1 & 1 & 1 & 1 & 1 \\ 
0.3 & 0.998 & 0.989 & 0.521 & 0.999 & 1 & 1 & 1 & 1 \\ 
0.5 & 1 & 0.998 & 0.989 & 0.992 & 1 & 1 & 1 & 1 \\ 
0.7 & 1 & 1 & 0.998 & 0.547 & 1 & 1 & 1 & 1 \\ 
0.9 & 1 & 1 & 0.998 & 0.864 & 1 & 1 & 1 & 1 \\ 
1 & 1 & 1 & 1 & 0.956 & 1 & 1 & 1 & 1 \\ 
1.5 & 1 & 1 & 1 & 0.998 & 1 & 1 & 1 & 1 \\ 
2 & 1 & 1 & 1 & 1 & 0.998 & 1 & 1 & 1 \\

\hline
\end{tabular}}
\end{table}

\section{Conclusion}
\label{Sec_Conclusion}

Logarithmic super divergence family acts as a super family of both LPD and LDPD family. 
Its usage in both statistical estimation and testing of hypotheses have been studied.
Along with the limitation of the first order influence function 
and the breakdown point under location model 
have also extensively studied. 
Computational exercises have shown that there exist a region of the parameter which 
usually performs better where outliers are present in the observations.


\begin{thebibliography}{99}
\bibitem{bhhj98}
Basu, A., I. R. Harris, N. L. Hjort, and M. C. Jones (1998). Robust and efficient 
estimation by minimising a density power divergence. {\em Biometrika}, {\bf 85}, 549--559.
\bibitem{be77}
Beran, R. J. (1977). Minimum Hellinger distance estimates for parametric 
models. {\em Annals of Statistics},  {\bf 5}, 445--463.
\bibitem{breg67}
Bregman, L. M. (1967). The relaxation method of finding the common point 
of convex sets and its application to the solution of problems in convex 
programming. {\em USSR Computational Mathematics and Mathematical Physics}, {\bf 7}, 
200--217. Original article is in {\em Zh. vychisl. Mat. mat. Fiz.}, {\bf 7}, pp. 620--631, 1967.
\bibitem{cr84}
Cressie, N. and T. R. C. Read (1984). Multinomial goodness-of-fit tests. {\em Journal 
of the Royal Statistical Society B}, {\bf 46}, 440--464.
\bibitem{cs63}
Csis\'{z}ar, I. (1963). Eine informations theoretische Ungleichung und ihre 
Anwendung auf den Beweis der Ergodizitat von Markoffschen Ketten. {\em Publ. Math. 
Inst. Hungar. Acad. Sci.}, {\bf 3}, 85--107.
\bibitem{feg08}
Fujisawa, H. and S. Eguchi. (2008). Robust parameter estimation with a small bias against heavy
contamination. {\em Journal of Multivariate Analysis}, {\bf 99}, 2053--2081.
\bibitem{fuji11}
Fujisawa, H. (2013). Normalized estimating equation for
robust parameter estimation. {\em Electronic Journal of Statistics}, {\bf 7}, 1587--1606. 
\bibitem{gmbp12}
Ghosh, A., I.R. Harris,  A. Maji, A. Basu, and L. Pardo (2013). The Robust Parametric Inference based on a
New Family of Generalized Density Power Divergence Measures. Technical Report, 
Bayesian and Interdisciplinary Research Unit, Indian Statistical Institute, India.
\bibitem{gmbp14a}
Ghosh, A., and A. Basu (2014). On Robustness of A Divergence based Test of Simple Statistical Hypothesis. 
Technical Report, Bayesian and Interdisciplinary Research Unit, Indian Statistical Institute, India.
\bibitem{hrrs86}
Hampel, F. R., E. Ronchetti, P. J. Rousseeuw, and W. Stahel (1986). 
{\em Robust Statistics: The Approach Based on Influence Functions}. New York, 
USA: John Wiley $\&$ Sons.
\bibitem{jhhb01}
Jones, M. C., N. L. Hjort, I. R. Harris, and A. Basu (2001). A comparison of 
related density-based minimum divergence estimators. {\em Biometrika}, {\bf 88}, 865--873.
\bibitem{kumbasu14}
Kumar, . and A. Basu (2014). Technical Report, 
Bayesian and Interdisciplinary Research Unit, Indian Statistical Institute, India.
\bibitem{lind94}
Lindsay, B. G. (1994). Efficiency versus robustness: The case for minimum 
Hellinger distance and related methods. {\em Annals of Statistics}, {\bf 22}, 1081--1114.
\bibitem{mgb14}
 Maji, A., A. Ghosh, and A. Basu (2014). 
The Logarithmic Super Divergence and
Asymptotic Inference Properties. 
Technical Report, 
Bayesian and Interdisciplinary Research Unit, Indian Statistical Institute, India.
\bibitem{mcb14}
 Maji, A., S. Chakraborty, and A. Basu (2014). Statistical Inference based on the
Logarithmic Power Divergence. Technical Report, 
Bayesian and Interdisciplinary Research Unit, Indian Statistical Institute, India.
\bibitem{pb04}
Park, C. and A. Basu (2004). Minimum disparity estimation: Asymptotic normality 
and breakdown point results. {\em Bulletin of Informatics and Cybernetics}, {\bf 36}, 19--33. 
Special Issue in Honor of Professor Takashi Yanagawa.
\bibitem{P00}
Pearson, K. (1900). On the criterion that a given system of deviations from the probable in
the case of a correlated system of variables is such that it can be reasonably supposed to
have arisen from random sampling. Philosophical Magazine, {\bf 50}, 157--175.
\bibitem{rny61}
Renyi, A. (1961). On measures of entropy and information. { \em In Proceedings
of Fourth Berkeley Symposium on Mathematical Statistics and Probability},
{\bf volume I}, pages 547--561. University of California.
\bibitem{simp87}
Simpson, D. G. (1987). Minimum Hellinger distance estimation for the 
analysis of count data. {\em Journal of the American Statistical 
Association}, {\bf 82}, 802--807.

\end{thebibliography}

\end{document}